\documentclass[12pt]{article}
\usepackage{amsmath}

\newcommand{\labela}[1]{\label{#1}}

\makeatletter
  
  \@addtoreset{equation}{section}
\makeatother

\title{Higher Derivative Correction to the Hawking Flux via Trace Anomaly}
\author{Tomoyoshi Hirata, Akihide  Shirasaka}

\begin{document}

\begin{titlepage}
\thispagestyle{empty}

KUNS-2135
\begin{flushright}
\end{flushright}

\bigskip

\begin{center}
\noindent{\Large \textbf{Higher Derivative Correction to the Hawking Flux via Trace Anomaly}}\\
\vspace{2cm} \noindent{ 
Akihide Shirasaka\footnote{e-mail:akihide@gauge.scphys.kyoto-u.ac.jp},}
Tomoyoshi Hirata\footnote{e-mail:hirata@gauge.scphys.kyoto-u.ac.jp}
\vspace{1cm}

 {\it  Department of Physics, Kyoto University, Kyoto 606-8502, Japan
 }

\vskip 2em
\end{center}

\begin{abstract}
In this paper we derive  Hawking radiation of black holes with higher derivative corrections by the method of trace anomaly.
Firstly we derive Hawking radiation for general spherical black holes.
We introduce a modified tortoise coordinate to it and find the analyticity of the coordinate.
Secondly we apply its method to a black hole with a higher derivative correction and derive the Hawking radiation of it.
We found that as an ordinary case the flux depends only on the surface gravity.
\end{abstract}

\end{titlepage}

\newpage

\section{Introduction}
Up to now, Hawking radiation \cite{Hawking:1974sw} has been investigated by several
methods. Hawking originally described the radiation by observing a
gravitational collapse, and Unruh \cite{Unruh:1976db} found there is a radiation even if
one considers
an eternal black hole. Following them, Christensen and Fulling \cite{Christensen:1977jc}
discovered new means which is based on a conformal anomaly near a horizon.
Recently, Robinson and Wilczek \cite{Robinson:2005pd} (a modified
discussion is given by \cite{Iso:2006wa} and \cite{Umetsu:2008cm}) showed that 
Hawking radiation can be derived by a gravitational anomaly, and many
investigations have been made in various
situations(\cite{Iso:2006ut,Murata:2006pt,Iso:2007hd},
\cite{Iso:2006xj}...\cite{Das:2007ru}).

In this paper, by CFT method based on a conformal anomaly, we survey the
energy momentum tensor in infinity which is originated from Hawking radiation in general spherical
symmetric black hole backgrounds. The general backgrounds, especially, contain  higher
derivative corrected black hole geometries which are important to know
 the effects from string theory. To study via CFT method, we must modify
 the definition of the tortoise coordinate in a spherical symmetric
 black hole which has been defined in the case of the metric having a
 property that the product of time and radial components is $-1$. And we
 shows the new definition of the tortoise coordinate is well-defined to prove
 the Kruskal metric has no coordinate singularity near the horizon.
Requiring there are no divergent physical quantities at the horizon in the Kruskal
geometry, we calculate the energy momentum tensor and find the result
agrees to that of \cite{Wu:2007sw,Vagenas:2006qb} in which they derive it by gravitational anomaly method. We review and simplify it in Appendix A.

This article is organized as follows. 

In section 2, preparing for CFT method to investigate the energy
momentum tensor of Hawking radiation, we suggest the new definition of
the tortoise coordinate and, in the Kruskal coordinate constructed from it, 
prove that there is no coordinate singularity near the horizon.

In section 3, applying the modified tortoise coordinate, we calculate
the energy momentum tensor of Hawking radiation in general spherical
black hole backgrounds.

In section 4, using the result of section 3, we derive the higher
derivative correction to Hawking flux in a specific case which action
gives a non-trivial property to the metric ($g_{tt}g_{rr}\neq -1$) 
\cite{Campanelli:1994jt} and
which would be the simplest one of the metric having such a property.

In Appendix A, we rederive the result of section 3 to use the
gravitational anomaly method.

In Appendix B, we present the evidence of the theory having the nature of CFT near the horizon.

\section{Kruskal coordinate of general spherical black holes}

In this section, for preparation of the later section we examine the feature of the Kruskal coordinate of the general spherical black holes. 

 In general $D$ dimensional spherical black holes, the metrics have a form of
 
\begin{align}
ds^2=e^{2A(r)}dt^2-e^{2B(r)}dr^2-r^2d\Omega^2_{D-2}. 
                                                       \labela{metric}
\end{align}

 Now we write
\begin{align}
e^{-2B(r)}=e^{2A(r)}L(r)^2,
                                        \labela{A(r),B(r),L(r)}
\end{align}
and define $r=r_+$ as the greatest real root of
\begin{align}
e^{2A(r)}=0,
                                       \labela{def of horizon}
\end{align}
which is the radial coordinate of the horizon. We suppose  the invariant volume
element
is non-singular and non-zero at the horizon and out of the horizon , then 
$L(r)$ becomes
non-singular and non-zero in this region, because we have
\begin{align}
\sqrt{-g}=L(r)^{-1}r^2\sin \theta.
                                           \labela{volume element}
\end{align}

In general $L(r)\neq1$, though we have $L(r)=1$ for ordinary Einstein-Hilbert action. So we must
modify the tortoise coordinate which is defined in the case of $L(r)=1$.
It is desirable that, at the horizon, the metric has no coordinate
singularity when we use the Kruskal coordinate constructed of the
modified tortoise coordinate.

Now we define the modified tortoise coordinate $r_*$ as the solution of
the differential equation,
\begin{align}
\frac{dr_*}{dr}=e^{B(r)-A(r)}.
                                          \labela{def of tortoise}
\end{align}

We will show that the above definition leads to no coordinate
singularities in the metric in the representation of the Kruskal
coordinate.

Using the Eddington-Finkelstein coordinate $u=t-r_*, v=t+r_*$ the
metric (\ref{metric}) can be written as
\begin{align}
ds^2=e^{2A(r)}dudv-r^2 d\Omega^2_{D-2}.
\labela{two dimensional metric}
\end{align}
To obtain an expression of a surface gravity $\kappa_+$
defined by the Killing equation
\begin{align}
\xi^{\mu}\nabla_\mu \xi^\nu=\kappa_+ \xi^\nu,   
\labela{killingequation}
\end{align}
where $\xi^\mu$ is a Killing vector, we  rewrite (\ref{two dimensional metric}) in the ingoing
Eddington-Finkelstein coordinate $(v,r,\Omega_{D-2})$. 
Then we have
\begin{align}
ds^2=e^{2A(r)}dv^2-2e^{A(r)+B(r)}dvdr-r^2 d\Omega^2_{D-2}
\labela{ingoingEF}
\end{align}
We find that $\xi^v$ survives and is a constant, and other components
vanish. By evaluating (\ref{killingequation}) near the horizon. we obtain
\begin{align}
\kappa_+=\frac{1}{2}L(r_+)(e^{2A(r)})'|_{r=r_+}. 
\labela{surface gravity}
\end{align}

We solve the equation (\ref{def of tortoise}) near the horizon. To do so 
we rewrite $e^{2A(r)}$ to
\begin{align}
e^{2A(r)}=\left(1-\frac{r_+}{r}\right)g(r),
\labela{A(r),g(r)}
\end{align}
where $g(r)$ is a holomorphic and non-zero function near $r=r_+$ (we
suppose the black hole is non-extremal). The l.h.s. of (\ref{def of
tortoise}) is
\begin{align}
e^{B(r)-A(r)}=\left(1+\frac{r_+}{r-r_+}\right)h(r),
\labela{B(r)-A(r)}
\end{align}
where $h(r)=L(r)^{-1}g(r)^{-1}$ which is holomorphic at the
horizon. Near the horizon, $h(r)$ has a Taylor expansion
\begin{align}
h(r)=a_0+a_1(r-r_+)+\frac{1}{2}a_2(r-r_+)^2+\dots,
\labela{Taylor}
\end{align}
where $a_0=h(r_+),a_1=h'(r_+)$, e.t.c.. So (\ref{B(r)-A(r)}) is
\begin{align}
(a_0+a_1 r_+)+\frac{a_0 r_+}{r-r_+}+(terms\ holomorphic\ at\ r=r_+),
\labela{tenkai}
\end{align}
and can be integrated to
\begin{align}
r_*=a_0 r_+ \ln \left(\frac{r}{r_+}-1\right)+p(r),
\labela{r_*motometa}
\end{align}
where $p(r)$ is a holomorphic function at $r=r_+$. The relation
between $a_0 r_+$ and $\kappa_+$ is
\begin{align}
a_0 r_+&=L(r)^{-1}g(r)^{-1}r|_{r=r_+} \nonumber \\
       &=[L(r_+)(e^{2A(r)})'|_{r=r_+}]^{-1}=\frac{1}{2\kappa_+}.
\labela{a_0r_+}
\end{align}
Finally we have
\begin{align}
r_*=\frac{1}{2\kappa_+}\ln \left(\frac{r}{r_+}-1\right)+p(r).
\labela{r_*kettei}
\end{align}
This is the similar result of the tortoise coordinate of the 
black hole obtained from the ordinary Einstein-Hilbert action.

Define the Kruskal coordinate,
\begin{align}
U=-e^{-\kappa_+ u}, V=e^{\kappa_+ v},
\labela{Kruskal}
\end{align}
using (\ref{r_*kettei}), the metric (\ref{two dimensional metric})
becomes
\begin{align}
ds^2=-\frac{r_+ e^{-p(r)}}{\kappa_+^2 r}dUdV-r^2 d\Omega^2_{D-2}.
\labela{kruskal nonsingular}
\end{align}
As there is no coordinate singularity at $r\ge r_+$, our definition (\ref{def of tortoise}) is reasonable.

Using the modified coordinate, we will see the Hawking flux via
CFT method and obtain the same result of the gravitational anomaly method.

\section{Method of Trace Anomaly of Two dimensional CFT}   \label{trace
anomaly method}
\
In \cite{Christensen:1977jc} it is reported that the Hawking flux of the Schwarzschild black hole is derived via the method of trace anomaly in two dimensional CFT.
And the method is applied to many kinds of black holes.
 For example the Hawking flux of  the four dimensional charged black hole without
higher derivative correction  was calculated in this method
\cite{Iso:2007hd}.

In this section, we apply the method to the general spherical black holes and  derive the Hawking flux of them.
As shown in Appendix B, near the horizon we obtain the
effective two dimensional theory with dilaton,  which metric is
\begin{align}
ds^2=e^{2A(r)}dt^2-e^{2B(r)}dr^2=e^{2A(r)}dudv         
                                  \labela{two dimensional metric}
\end{align}

The energy-momentum conservation and the current
conservation are
\begin{align}
\nabla_\mu T^\mu_{\ \nu}&=F_{\mu\nu}J^\nu, \labela{cft em conservation}\\
\nabla_\mu J^\mu&=0.  \labela{cft current conservation}
\end{align}
And there is matter fields which contribute to the Hawking radiation,
the energy-momentum tensor has the trace anomaly
\begin{align}
T^\mu_{\ \mu}=\frac{1}{24\pi}R.
\labela{trace anomaly}
\end{align} 
The two dimensional chiral current
$J^{5\mu}=\frac{\epsilon^{\mu\nu}}{\sqrt{-g}}J_\nu$ satisfies the anomalous
conservation law
\begin{align}
\nabla_\mu
 J^{5\mu}=\frac{e^2}{2\pi}\frac{\epsilon^{\mu\nu}}{\sqrt{-g}}F_{\mu\nu}.
\labela{chiral current anomaly}
\end{align}
From (\ref{cft current conservation}) and (\ref{chiral current anomaly}), we have
\begin{align}
\partial_u J_v+\partial_v J_u&=0,
\labela{cft current conservation new}\\
\partial_u J_v-\partial_v J_u&=\frac{e^2}{\pi}F_{uv}.
\labela{chiral current anomaly new}
\end{align}
Under the Lorentz gauge $\nabla^\mu A_\mu=0$, these equations can be solved as
\begin{align}
J_u&=j(u)+\frac{e^2}{\pi}A_u(r),
\labela{Ju solved}\\
J_v&=\tilde{j}(v)+\frac{e^2}{\pi}A_v(r),
\labela{Jv solved}
\end{align}
where $j(u)$ ($\tilde{j}(v)$) is a holomorphic (an anti-holomorphic)
function which we will determine later.

In the Kruskal coordinate (\ref{Kruskal})
we have 
\begin{align}
J_U=-\frac{1}{\kappa_+ U}\left[j(u)+\frac{e^2}{\pi}A_u(r)\right].
\labela{JU no shiki}
\end{align}
As there is no singularity on the horizon when we use the Kruskal coordinate,  we require that $J_U$ has a finite value on the
horizon, so we have
\begin{align}
j(u)|_{r\rightarrow r_+}=-\frac{e^2}{\pi}A_u(r_+).
\labela{ju horizon}
\end{align}
And we require there is no ingoing mode:
\begin{align}
\tilde{j}(v)|_{r\rightarrow \infty}=0.
\labela{tilde jv infinity}
\end{align}
With these requirements we can calculate the current at infinity
\begin{align}
J^r=\frac{dr}{dr_*}J^{r_*}\rightarrow -\frac{e^2}{\pi}A_u(r_+)\ (as\ r\rightarrow\infty).
\labela{cft Jr infinity}
\end{align}

Next, we calculate the energy momentum tensor. From (\ref{cft em
conservation}) and (\ref{trace anomaly}) with the Lorentz gauge $\nabla^\mu
A_\mu=0$, we find
\begin{align}
T_{uu}&=t(u)+\frac{1}{12\pi}(\partial_u^2 A(r)-(\partial_u
 A(r))^2)+\frac{e^2}{\pi}A_u(r)^2+2A_u(r) j(u),
\labela{Tuu no shiki}\\
T_{vv}&=\tilde{t}(v)+\frac{1}{12\pi}(\partial_v^2 A(r)-(\partial_v
 A(r))^2)
+\frac{e^2}{\pi}A_v(r)^2+2A_v(r) j(u),
\labela{Tvv no shiki}
\end{align}
here $t(u)$ and  $\tilde{t}(v)$ are a holomorphic and an anti-holomorphic
function respectively.

As above, we require that $T_{UU}\sim\frac{1}{U^2}T_{uu}$ must be
regular on the future horizon. Then we have
\begin{align}
t(u)|_{r\rightarrow r_+}&=-\frac{1}{12\pi}(\partial_u^2 A(r)-(\partial_u
 A(r))^2)|_{r=r_+}+\frac{e^2}{\pi}A_u(r_+)^2\\
&=\frac{1}{192\pi}[(e^{-2B(r)})^{'}(e^{2A(r)})^{'}]_{r=r_+}+\frac{e^2}{\pi}A_u(r_+)^2.
\labela{tu decided}
\end{align}
We also require that there is no ingoing flux from infinity
\begin{align}
\tilde{t}(v)|_{r\rightarrow\infty}=0.
\labela{tv decided}
\end{align}

Then we obtain the energy-momentum tensor:
\begin{align}
T^r_t=\frac{dr}{dr_*}T^{r_*}_t
&\rightarrow\frac{1}{192\pi}[(e^{-2B(r)})^{'}(e^{2A(r)})^{'}]_{r=r_+}+\frac{e^2}{\pi}A_u(r_+)^2 \nonumber\\
&=\frac{1}{48\pi}\kappa_+^2+\frac{e^2}{\pi}A_u(r_+)^2.
\labela{cft hawking decided}
\end{align}

The first term of (\ref{cft hawking decided}) is derived from the Hawking flux and the second one is derived from electric-magnetic potential of black holes.
We see that the flux of Hawking radiation is dominated by the surface gravity $\kappa_+$ in any cases.

In \cite{Wu:2007sw,Vagenas:2006qb}, using the gravitational anomaly method, they also derive the Hawking flux of the general spherical black holes. We summarize their derivation in  appendix A.

\section{Higher Derivative Correction of the Hawking Flux}

In this section, by using the method we established in the last section  we calculate the Hawking flux from the four dimensional charged black hole
which includes the higher derivative
correction.
We set the action  
 \cite{Campanelli:1994jt} as follows
\begin{align}
I=\frac{1}{16\pi G}\int d^4 x \sqrt{-g} \left[ R + F_{\mu\nu} F^{\mu\nu} + a
 R^2 +b R_{\mu\nu} R^{\mu\nu} +c R_{\mu\nu\rho\sigma}
 R^{\mu\nu\rho\sigma} \right],   
                                            \labela{action}
\end{align}
where $G$ is the four dimensional Newton constant and $a,b$ and $c$ are
infinitesimal parameters with the dimension $(mass)^2$.
The corrected metric of the charged black hole is
\begin{align}
ds^2=e^{2A(r)}dt^2-e^{2B(r)}dr^2-r^2 d\Omega_2^2.
                                            \labela{metric2d}
\end{align}
where up to order one of $a$, $b$ and $c$
\begin{align}
e^{2A(r)}=1-\frac{2M}{r}+\frac{Q^2}{r^2}+(b+4c)\left(-\frac{2Q^2}{r^4}+\frac{2Q^2M}{r^5}-\frac{2Q^4}{5r^6}\right),
                                            \labela{exp[2A]}
\end{align}
\begin{align}
e^{-2B(r)}=1-\frac{2M}{r}+\frac{Q^2}{r^2}+(b+4c)\left(-\frac{4Q^2}{r^4}+\frac{6Q^2M}{r^5}-\frac{12Q^4}{5r^6}\right),         
                                           \labela{exp[-2B]}
\end{align}
$M$ and $Q$ is the mass and the charge of the uncorrected black hole.
And the corrected electro-magnetic field is
\begin{align}
A_t(r)=\frac{Q}{r}+\frac{Q^3}{5r^5}(b+4c).
                                           \labela{A_t(r)}
\end{align}

The uncorrected black hole have a property 
\begin{align}
A(r)=-B(r),
                                           \labela{A(r)}
\end{align}
however, this property is corrected.
Up to order one of $a$, $b$ and $c$, we have
\begin{align}
e^{-2B(r)}=e^{2A(r)}L(r)^2,\  where\  L(r)^2=1-\frac{2Q^2}{r^4}(b+4c).
                                           \labela{relation}
\end{align}
The exact form of the $L(r)$ is shown in \cite{Campanelli:1994jt}.

We know that the only quantity which is needed to obtain the Hawking flux is the surface gravity $\kappa_+$ (\ref{surface gravity}).
Now we can calculate the higher derivative correction to the Hawking flux.
From (\ref{exp[2A]}) and (\ref{exp[-2B]}), up to order one of $a$, $b$ and $c$, the Hawking flux is
\begin{eqnarray}
\frac{\kappa_+^2}{48\pi}= F_o+ \delta F
\end{eqnarray}
where $F_0$ is an original Hawking flux  
\begin{equation}
F_0=\frac{(M^2-Q^2 +M \sqrt{M^2-Q^2})^2}{48\pi(M+\sqrt{M^2-Q^2})^6}
\end{equation}
and $\delta F$ is the higher derivative correction of it
\begin{equation}
\delta F=(b+4c)\frac{20M^4Q^2 -25M^2Q^4 +6Q^6+ \sqrt{M^2-Q^2}(20M^3Q^4-15MQ^6) }{240\pi(M+\sqrt{M^2-Q^2})^{10}}    .                         
\labela{kappa_+^2}
\end{equation}

\section{Conclusion and Outlook}
In this paper we obtain an explicit formula for the higher derivative corrected Hawking flux by the CFT method.
As a by-product we obtain the general formula for any kind of spherical black holes in any kind of modified gravity by the CFT method. 
We found that the form (\ref{cft hawking decided}) is the same as an ordinary black hole, and only depends on 
the surface gravity $\kappa_+$.
We saw that tortoise coordinate $r_*$ defined as (\ref{def of tortoise}) has no singularity in the region $r\ge r_+$.
In  Appendix \ref{sec grav}, we simplified the gravitational anomaly method given by \cite{Robinson:2005pd} and many other papers.

As a future work, it is  interesting  to apply this
method to non-spherical black holes. To find the explicit relationship
between two methods \cite{Christensen:1977jc} and \cite{Robinson:2005pd} is also an interesting question remained.

\vskip3mm

  \noindent{\bf Acknowledgement}
  
 We are extremely grateful to T.Morita S.Yamato  for valuable discussions.

The work of T.H is supported in part by Japan Society for the Promotion of Science Research Fellowships for 
 Young Scientists.

\vskip2mm

\appendix
\section{Calculation of Corrected Hawking Flux via Gravitational Anomaly 
}\label{sec grav}\ 
In this appendix, we calculate the Hawking flux from four dimensional 
black hole   with higher derivative correction  by using the method studied by 
Robinson and Wilczek \cite{Robinson:2005pd} (a modified derivation 
was shown by \cite{Iso:2006wa} and \cite{Umetsu:2008cm}). These papers are based on a gravitational 
anomaly of an effective chiral two dimensional theory. Because we treat 
with charged black hole, we must know $U(1)$ current of the two 
dimensional theory (as discussed in \cite{Iso:2006wa}). 
Since the theory is described as the two dimensional CFT in the near horizon region (Appendix B),
we can separate ingoing modes and outgoing modes between the horizon $r_+$ 
and slightly outside  $r_+ + \epsilon$ . We will call the region 
``shell region''.
We decompose the two dimensional $U(1)$ current as 
\begin{align} 
J^\mu(r)=J_{(o)}^\mu(r)\Theta_+(r)+(J_{(H)}^\mu(r)+K_{(H)}^\mu(r))H(r), 
\labela{decompose current} 
\end{align} 
where $\Theta_+(r)=\Theta(r-\epsilon)$ and $H(r)=1-\Theta_+(r)$
($\Theta(r)=0,1$ for $r< 0,r>0$ respectively), these are  step functions which take values in the shell region and out the shell region respectively.  $J_{(H)}^\mu(r)$ and $K_{(H)}^\mu(r)$ are respectively outgoing current and ingoing current.
The current $J_{(H)}^\mu$ and  $K_{(H)}^\mu$ obey conservation laws with 
consistent anomaly.
\begin{align} 
\nabla_\mu J^\mu_{(H)}=\frac{e^2}{4\pi\sqrt{-g}}\partial_r A_t 
\labela{consistent anomaly} 
\end{align} 
and 
\begin{align} 
\nabla_\mu K^\mu_{(H)}=-\frac{e^2}{4\pi\sqrt{-g}}\partial_r A_t. 
\labela{consistent anomaly ingoing} 
\end{align} 

These equations can be integrated to 
\begin{align} 
e^{A(r)+B(r)}J_{(H)}^r 
(r)=e^{A(r_+)+B(r_+)}c_H+\frac{e^2}{4\pi}(A_t(r)-A_t(r_+)), 
\labela{inside current} 
\end{align} 
and 
\begin{align} 
e^{A(r)+B(r)}K_{(H)}^r 
(r)=-\frac{e^2}{4\pi}(A_t(r)), 
\labela{inside current ingoing} 
\end{align} 
where $c_H$ is an integration constant. We set the integral constant of
$K_{(H)}^r$  to zero, which condition means there is no ingoing mode from the infinity.

Outside the shell region, $J_{(o)}^\mu$ satisfies an ordinary conservation 
\begin{align} 
\nabla_\mu J^\mu_{(o)}=0, 
\labela{ordinary conservation} 
\end{align} 
and 
\begin{align} 
e^{A(r)+B(r)}J_{(o)}^r=c_o, 
\labela{outside current} 
\end{align} 
where $c_o$ is an integration constant which represents the value of the 
$U(1)$ current in infinity.

 As \cite{Iso:2006wa} we
introduce a new current 
\begin{align}
e^{A(r)+B(r)}\tilde{J}_{(H)}^\mu&=e^{A(r)+B(r)}J_{(H)}^\mu+\frac{e^2}{4\pi}A_t(r), 
\labela{tilde J(H)}
\end{align}
$\tilde{J}_{(H)}^\mu$ satisfies the conservation law with covariant anomaly
\begin{align}
\nabla_\mu
 \tilde{J}^\mu_{(H)}=\frac{e^2}{2\pi\sqrt{-g}}\partial_r
 A_t.
\labela{covariant anomaly}
\end{align}

We  require that $J^\mu(r)$ is smooth at the boundary of the shell and  on the horizon the covariant current  vanishes 
\cite{Robinson:2005pd}\cite{Iso:2006wa}.
By these requirements we obtain
\begin{align}
c_o=e^{A(r_+)+B(r_+)}c_H-\frac{e^2}{4\pi}A_t(r_+).\labela{int const rel}
\end{align} 
and
\begin{align}
e^{A(r_+)+B(r_+)}c_H=-\frac{e^2}{4\pi}A_t(r_+).
\labela{cH decided}
\end{align}
From (\ref{int const rel}) and (\ref{cH decided}), we conclude
\begin{align}
c_o=-\frac{e^2}{2\pi}A_t(r_+).
\labela{cO decided}
\end{align}

Next we calculate energy-momentum tensor.
As we treat with $J^\mu$ we separate the energy-momentum tensor as
\begin{align} 
T^\mu_{\ \nu}(r)=T_{\ \nu(o)}^\mu(r)\Theta_+(r)+(T_{\ \nu(H)}^\mu(r)+S_{\ \nu(H)}^\mu(r))H(r),
\labela{decompose energy} 
\end{align} 
where $T_{\ \nu(H)}^\mu$ is the outgoing current in the shell region and $S_{\ \nu(H)}^\mu(r)$ is the ingoing one there.

These currents satisfy the following conservation laws.
\begin{eqnarray}
\nabla_\mu T_{\ \nu(H)}^\mu(r)&=&F_{\mu\nu}J^\mu_{(H)}+A_\nu(\nabla_\mu
 J^\mu_{(H)})+\mathcal{A}_\nu\\
\nabla_\mu S_{\ \nu(H)}^\mu(r)&=&F_{\mu\nu}K^\mu_{(H)}+A_\nu(\nabla_\mu
 K^\mu_{(H)})-\mathcal{A}_\nu\\
  \nabla_\mu T_{\ \nu(o)}^\mu(r)&=&F_{\mu\nu}J^\mu_{(o)}+A_\nu(\nabla_\mu
 J^\mu_{(o)})
 \labela{energy momentum conservation}
\end{eqnarray}
where 
\begin{align}
\mathcal{A}_\nu=\frac{1}{96\pi\sqrt{-g}}\epsilon^{\beta\delta}\partial_{\delta}\partial_{\alpha}\Gamma_{\nu\beta}^{\alpha}
\labela{consistent gravitational}
\end{align}
is the consistent gravitational anomaly. In the metric (\ref{metric2d}) we can calculate
\begin{align}
e^{A(r)+B(r)}\mathcal{A}_r&=0,
                       \labela{mathcal A r}\\
e^{A(r)+B(r)}\mathcal{A}_t&=\partial_r N_t^r,
                       \labela{mathcal A t}\\
&\left(N_t^r=\frac{1}{192\pi}(e^{-2B(r)}(e^{2A(r)})^{'})^{'}\right).
                       \labela{N}
\end{align}
Outside the shell region, from (\ref{energy momentum conservation})
\begin{align}
e^{A(r)+B(r)}T^r_{\ t(o)}=a_o+c_oA_t(r),
\labela{outside flux}
\end{align}
where $a_o$ is the integration constant which means the energy momentum flux
at infinity.

In the shell region, from (\ref{energy momentum conservation}) we have
\begin{eqnarray}
e^{A(r)+B(r)}T^r_{\ t(H)}&=&a_H+\left[c_oA_t(r)+
\frac{e^2}{4\pi}A_t^2+N_t^r\right]^r_{r_H},
\labela{inside flux1}\\
e^{A(r)+B(r)}K^r_{\ t(H)}&=&-\left(c_oA_t(r)+
\frac{e^2}{4\pi}A_t^2+N_t^r\right),
\labela{inside flux2}
\end{eqnarray}
where $a_H$ is an integration constant, and we again set the integral
constant of $K^r_{\ t}$ to zero.

From the smoothness of the $T^{\mu}_{\ \nu}$, we have
\begin{align}
a_o=a_H+\frac{e^2}{4\pi}A_t(r_+)^2-N_t^r(r_+).
\labela{ao ah rel}
\end{align}

To obtain $a_o$, we have to know $a_H$. We take the
similar argument of evaluating of $c_H$.
 
We introduce a new energy momentum tensor:
\begin{align}
e^{A(r)+B(r)}\tilde{T}^r_{\ t}=e^{A(r)+B(r)}T^r_{\
 t}+\frac{1}{96\pi}e^{2A(r)-2B(r)}(A^{''}(r)-2A^{'}(r)^2).
\labela{cov and cons rel}
\end{align}
The new energy momentum tensor satisfies a conservation law with a covariant anomaly,
\begin{equation}
\nabla_\mu \tilde{T}_{\ \nu(H)}^\mu(r)=F_{\mu\nu}J^\mu_{(H)}+A_\nu(\nabla_\mu
 J^\mu_{(H)})+\tilde{\mathcal{A}}_\nu,
\end{equation}
where
\begin{align}
 \tilde{\mathcal{A}}_\nu \equiv\frac{1}{96\pi\sqrt{-g}}\epsilon_{\mu\nu}\partial^\mu R
\labela{covariant gravitational anomaly}
\end{align}

 Now we require that the covariant flux must vanish at the horizon. This requirement leads to
\begin{align}
a_H=2N_t^r(r_+),
\end{align}
and from (\ref{ao ah rel})
\begin{align}
a_o=\frac{e^2}{4\pi}A_t(r_+)^2+N_t^r(r_+).
\labela{ao decided}
\end{align}
 Using of the surface gravity $\kappa_+$ (\ref{surface gravity}),
\begin{align}
a_o=\frac{e^2}{4\pi}A_t(r_+)^2+\frac{\kappa_+^2}{48\pi}.
\labela{ao final}
\end{align}
This is the Hawking flux at infinity and is the same one as (\ref{cft hawking decided})

\section{The Emergence of the Two dimensional CFT Near the Horizon}
Now we describe the reason of having a property of two dimensional  CFT
near the horizon. For simplicity, we treat with a
real scalar field $\phi$ on the general four dimensional spherical
symmetric black hole spacetime which metric is given by (\ref{metric}).

First, we see the kinetic term describes an effective two dimensional
theory near the horizon. The kinetic term is given by
\begin{align}
I_k&=\int d^4 x \sqrt{-g}\ \frac{1}{2}g^{\mu\nu}\partial_\mu \phi
 \partial_\nu\phi \nonumber\\
&=\int dtdrd\theta d\varphi\ r^2\sin\theta \left(\frac{1}{2}e^{A(r)+B(r)}g^{\mu\nu}\partial_\mu \phi
 \partial_\nu\phi \right).
\labela{kinetic term}
\end{align}
Under a transformation $r\rightarrow r_*$ ($r_*$ is defined by (\ref{def
of tortoise})), this becomes
\begin{align}
I_k=\int dt dr_* d\theta d\varphi\ r^2\sin\theta
 \frac{1}{2}\left[(\partial_t\phi)^2-(\partial_{r_*}\phi)^2-e^{2A(r)}r^{-2}\left((\partial_\theta\phi)^2+(\partial_\varphi
 \phi)^2\right)\right].
\labela{r_* effective}
\end{align}
Thus, as $r\rightarrow r_+$, only first two terms survive and $I_k$ would describe a
two dimensional theory. More explicitly, a partial wave
expansion
\begin{align}
\phi=\sum_{l,m}\phi_{lm}(t,r_*)Y_{lm}(\theta,\varphi),
\labela{partial wave}
\end{align}
where $Y_{lm}(\theta,\varphi)$ are spherical harmonics, yields
\begin{align}
I_k=\int dt dr_* r^2 \sum_{l,m} \frac{1}{2}[|\partial_t
 \phi_{lm}(t,r_*)|^2-|\partial_{r_*} \phi_{lm}(t,r_*)|^2],
\labela{partial wave action}
\end{align}
near the horizon. So the theory becomes a two dimensional theory with
dilaton.

Next, we show vanishing of the mass term and interaction terms. It is very
easy to show. We consider $\phi^n\ (n=2,3,\dots)$ terms, and
\begin{align}
\int dt dr d\theta d\varphi\ \phi^n=\int dt dr_* d\theta d\varphi\ r^2
 \sin\theta L(r)e^{2A(r)}\phi^n.
\end{align}
So the mass term and interaction terms vanish near the horizon.

Thus we conclude that near the horizon the theory becomes an effective two
dimensional CFT.


\begin{thebibliography}{99}
%\cite{Campanelli:1994jt}
\bibitem{Campanelli:1994jt}
  M.~Campanelli, C.~O.~Lousto and J.~Audretsch,
  ``Perturbative metric of charged black holes in quadratic gravity,''
  Phys.\ Rev.\  D {\bf 51}, 6810 (1995)
  [arXiv:gr-qc/9412001].
  %%CITATION = PHRVA,D51,6810;%%

%\cite{Robinson:2005pd}
\bibitem{Robinson:2005pd}
  S.~P.~Robinson and F.~Wilczek,
  ``A relationship between Hawking radiation and gravitational anomalies,''
  Phys.\ Rev.\ Lett.\  {\bf 95}, 011303 (2005)
  [arXiv:gr-qc/0502074].
  %%CITATION = PRLTA,95,011303;%%

%\cite{Iso:2006wa}
\bibitem{Iso:2006wa}
  S.~Iso, H.~Umetsu and F.~Wilczek,
  ``Hawking radiation from charged black holes via gauge and gravitational
  anomalies,''
  Phys.\ Rev.\ Lett.\  {\bf 96}, 151302 (2006)
  [arXiv:hep-th/0602146].
  %%CITATION = PRLTA,96,151302;%%

%\cite{Iso:2006ut}
\bibitem{Iso:2006ut}
  S.~Iso, H.~Umetsu and F.~Wilczek,
  ``Anomalies, Hawking radiations and regularity in rotating black holes,''
  Phys.\ Rev.\  D {\bf 74}, 044017 (2006)
  [arXiv:hep-th/0606018].
  %%CITATION = PHRVA,D74,044017;%%

%\cite{Murata:2006pt}
\bibitem{Murata:2006pt}
  K.~Murata and J.~Soda,
  ``Hawking radiation from rotating black holes and gravitational  anomalies,''
  Phys.\ Rev.\  D {\bf 74}, 044018 (2006)
  [arXiv:hep-th/0606069].
  %%CITATION = PHRVA,D74,044018;%%

%\cite{Christensen:1977jc}
\bibitem{Christensen:1977jc}
  S.~M.~Christensen and S.~A.~Fulling,
  ``Trace Anomalies And The Hawking Effect,''
  Phys.\ Rev.\  D {\bf 15}, 2088 (1977).
  %%CITATION = PHRVA,D15,2088;%%

%\cite{Iso:2007hd}
\bibitem{Iso:2007hd}
  S.~Iso, T.~Morita and H.~Umetsu,
  ``Fluxes of Higher-spin Currents and Hawking Radiations from Charged Black
  Holes,''
  arXiv:0705.3494 [hep-th].
  %%CITATION = ARXIV:0705.3494;%%

%\cite{Hawking:1974sw}
\bibitem{Hawking:1974sw}
  S.~W.~Hawking,
  ``Particle Creation By Black Holes,''
  Commun.\ Math.\ Phys.\  {\bf 43}, 199 (1975)
  [Erratum-ibid.\  {\bf 46}, 206 (1976)].
  %%CITATION = CMPHA,43,199;%%

%\cite{Unruh:1976db}
\bibitem{Unruh:1976db}
  W.~G.~Unruh,
  ``Notes on black hole evaporation,''
  Phys.\ Rev.\  D {\bf 14}, 870 (1976).
  %%CITATION = PHRVA,D14,870;%%

%\cite{Wu:2007sw}
\bibitem{Wu:2007sw}
  S.~Q.~Wu and J.~J.~Peng,
  ``Hawking radiation from the Reissner-Nordstr\'{o}m black hole with a
  global monopole via gravitational and gauge anomalies,''
  Class.\ Quant.\ Grav.\  {\bf 24}, 5123 (2007)
  [arXiv:0706.0983 [hep-th]].
  %%CITATION = CQGRD,24,5123;%%

%\cite{Iso:2006xj}
\bibitem{Iso:2006xj}
  S.~Iso, T.~Morita and H.~Umetsu,
   ``Quantum anomalies at horizon and Hawking radiations in Myers-Perry black
  holes,''
  JHEP {\bf 0704}, 068 (2007)
  [arXiv:hep-th/0612286].
  %%CITATION = JHEPA,0704,068;%%

%\cite{Iso:2007kt}
\bibitem{Iso:2007kt}
  S.~Iso, T.~Morita and H.~Umetsu,
  ``Higher-spin currents and thermal flux from Hawking radiation,''
  Phys.\ Rev.\  D {\bf 75}, 124004 (2007)
  [arXiv:hep-th/0701272].
  %%CITATION = PHRVA,D75,124004;%%

%\cite{Iso:2007nc}
\bibitem{Iso:2007nc}
  S.~Iso, T.~Morita and H.~Umetsu,
  ``Higher-spin Gauge and Trace Anomalies in Two-dimensional Backgrounds,''
  arXiv:0710.0453 [hep-th].
  %%CITATION = ARXIV:0710.0453;%%

%\cite{Iso:2007nf}
\bibitem{Iso:2007nf}
  S.~Iso, T.~Morita and H.~Umetsu,
  ``Hawking Radiation via Higher-spin Gauge Anomalies,''
  Phys.\ Rev.\  D {\bf 77}, 045007 (2008)
  [arXiv:0710.0456 [hep-th]].
  %%CITATION = PHRVA,D77,045007;%%

%\cite{Xu:2006tq}
\bibitem{Xu:2006tq}
  Z.~Xu and B.~Chen,
  ``Hawking radiation from general Kerr-(anti)de Sitter black holes,''
  Phys.\ Rev.\  D {\bf 75}, 024041 (2007)
  [arXiv:hep-th/0612261].
  %%CITATION = PHRVA,D75,024041;%%

%\cite{Banerjee:2007qs}
\bibitem{Banerjee:2007qs}
  R.~Banerjee and S.~Kulkarni,
  ``Hawking Radiation and Covariant Anomalies,''
  Phys.\ Rev.\  D {\bf 77}, 024018 (2008)
  [arXiv:0707.2449 [hep-th]].
  %%CITATION = PHRVA,D77,024018;%%

%\cite{Gangopadhyay:2007fm}
\bibitem{Gangopadhyay:2007fm}
  S.~Gangopadhyay and S.~Kulkarni,
   ``Hawking radiation in GHS and non-extremal D1-D5 blackhole via covariant
  anomalies,''
  Phys.\ Rev.\  D {\bf 77}, 024038 (2008)
  [arXiv:0710.0974 [hep-th]].
  %%CITATION = PHRVA,D77,024038;%%

%\cite{Banerjee:2007uc}
\bibitem{Banerjee:2007uc}
  R.~Banerjee and S.~Kulkarni,
  ``Hawking Radiation, Effective Actions and Covariant Boundary Conditions,''
  Phys.\ Lett.\  B {\bf 659}, 827 (2008)
  [arXiv:0709.3916 [hep-th]].
  %%CITATION = PHLTA,B659,827;%%

%\cite{Iso:2008sq}
\bibitem{Iso:2008sq}
  S.~Iso,
  ``Hawking Radiation, Gravitational Anomaly and Conformal Symmetry - the
  Origin of Universality -,''
  arXiv:0804.0652 [hep-th].
  %%CITATION = ARXIV:0804.0652;%%

%\cite{Wu:2008yx}
\bibitem{Wu:2008yx}
  X.~n.~Wu, C.~G.~Huang and J.~R.~Sun,
  ``On Gravitational anomaly and Hawking radiation near weakly isolated
  horizon,''
  arXiv:0801.1347 [gr-qc].
  %%CITATION = ARXIV:0801.1347;%%

%\cite{Ma:2007xr}
\bibitem{Ma:2007xr}
  Z.~Z.~Ma,
  ``Hawking radiation of black p-branes via gauge and gravitational
  anomalies,''
  arXiv:0709.3684 [hep-th].
  %%CITATION = ARXIV:0709.3684;%%

%\cite{Kim:2007gj}
\bibitem{Kim:2007gj}
  W.~Kim and J.~J.~Oh,
  ``Greybody Factor and Hawking Radiation of Charged Dilatonic Black Holes,''
  arXiv:0709.1754 [hep-th].
  %%CITATION = ARXIV:0709.1754;%%


%\cite{Gangopadhyay:2007hr}
\bibitem{Gangopadhyay:2007hr}
  S.~Gangopadhyay,
  ``Hawking radiation in GHS blackhole, Effective action and Covariant Boundary
  condition,''
  Phys.\ Rev.\  D {\bf 77}, 064027 (2008)
  [arXiv:0712.3095 [hep-th]].
  %%CITATION = PHRVA,D77,064027;%%

%\cite{Gangopadhyay:2008zw}
\bibitem{Gangopadhyay:2008zw}
  S.~Gangopadhyay,
  ``Hawking radiation in Reissner-Nordstr\'{o}m blackhole with a global
  monopole via Covariant anomalies and Effective action,''
  arXiv:0803.3492 [hep-th].
  %%CITATION = ARXIV:0803.3492;%%





%\cite{Peng:2008ru}
\bibitem{Peng:2008ru}
  J.~J.~Peng and S.~Q.~Wu,
  ``Covariant anomalies and Hawking radiation from charged rotating black
  strings in anti-de Sitter spacetimes,''
  Phys.\ Lett.\  B {\bf 661}, 300 (2008)
  [arXiv:0801.0185 [hep-th]].
  %%CITATION = PHLTA,B661,300;%%

%\cite{Huang:2007ed}
\bibitem{Huang:2007ed}
  C.~G.~Huang, J.~R.~Sun, X.~n.~Wu and H.~Q.~Zhang,
  ``Gravitational Anomaly and Hawking Radiation of Brane World Black Holes,''
  arXiv:0710.4766 [hep-th].
  %%CITATION = ARXIV:0710.4766;%%

%\cite{Murata:2007zr}
\bibitem{Murata:2007zr}
  K.~Murata and U.~Miyamoto,
  ``Hawking radiation of a vector field and gravitational anomalies,''
  Phys.\ Rev.\  D {\bf 76}, 084038 (2007)
  [arXiv:0707.0168 [hep-th]].
  %%CITATION = PHRVA,D76,084038;%%

%\cite{Kim:2007ge}
\bibitem{Kim:2007ge}
  W.~Kim and H.~Shin,
  ``Anomaly Analysis of Hawking Radiation from Acoustic Black Hole,''
  JHEP {\bf 0707}, 070 (2007)
  [arXiv:0706.3563 [hep-th]].
  %%CITATION = JHEPA,0707,070;%%

%\cite{Yu:2007qu}
\bibitem{Yu:2007qu}
  H.~Yu and W.~Zhou,
  ``Relationship between Hawking radiation from black holes and spontaneous
  excitation of atoms,''
  Phys.\ Rev.\  D {\bf 76}, 027503 (2007)
  [arXiv:0706.2207 [hep-th]].
  %%CITATION = PHRVA,D76,027503;%%

%\cite{Jiang:2007pe}
\bibitem{Jiang:2007pe}
  Q.~Q.~Jiang, S.~Q.~Wu and X.~Cai,
  ``Anomalies and de Sitter radiation from the generic black holes in de
  Sitter spaces,''
  Phys.\ Lett.\  B {\bf 651}, 65 (2007)
  [arXiv:0705.3871 [hep-th]].
  %%CITATION = PHLTA,B651,65;%%

%\cite{Miyamoto:2007ue}
\bibitem{Miyamoto:2007ue}
  U.~Miyamoto and K.~Murata,
  ``On Hawking radiation from black rings,''
  Phys.\ Rev.\  D {\bf 77}, 024020 (2008)
  [arXiv:0705.3150 [hep-th]].
  %%CITATION = PHRVA,D77,024020;%%

%\cite{Chen:2007pp}
\bibitem{Chen:2007pp}
  B.~Chen and W.~He,
  ``Hawking Radiation of Black Rings from Anomalies,''
  arXiv:0705.2984 [gr-qc].
  %%CITATION = ARXIV:0705.2984;%%

%\cite{Vagenas:2006qb}
\bibitem{Vagenas:2006qb}
  E.~C.~Vagenas and S.~Das,
  ``Gravitational anomalies, Hawking radiation, and spherically symmetric
  black holes,''
  JHEP {\bf 0610}, 025 (2006)
  [arXiv:hep-th/0606077].
  %%CITATION = JHEPA,0610,025;%%

%\cite{Das:2007ru}
\bibitem{Das:2007ru}
  S.~Das, S.~P.~Robinson and E.~C.~Vagenas,
  ``Gravitational anomalies: a recipe for Hawking radiation,''
  arXiv:0705.2233 [hep-th].
  %%CITATION = ARXIV:0705.2233;%%

%\cite{Umetsu:2008cm}
\bibitem{Umetsu:2008cm}
  K.~Umetsu,
  ``Ward Identities in the derivation of Hawking radiation from Anomalies,''
  arXiv:0804.0963 [hep-th].
  %%CITATION = ARXIV:0804.0963;%%














\end{thebibliography}
\end{document}